\documentclass[12pt,a4paper]{article}
\usepackage{ifpdf}
\ifpdf
\pdfpageattr {/Group << /S /Transparency /I true /CS /DeviceRGB>>}
\fi
\usepackage{jheppub}
\usepackage{amsmath}
\usepackage{caption}
\usepackage{subcaption}

\newcommand{\beq}{\begin{equation}}
\newcommand{\eeq}{\end{equation}}
\newcommand{\bea}{\begin{eqnarray}}
\newcommand{\eea}{\end{eqnarray}}

\newcommand{\eps}{\epsilon}

\newcommand{\phin}[1]{\phi_{(#1)}(t,\tau,x)}

\begin{document}

\title{AdS (In)stability: Lessons From The Scalar Field}
\author[a]{Pallab Basu,} \author[b]{Chethan Krishnan,} \author[b]{P. N. Bala Subramanian}
\emailAdd{pallabbasu@gmail.com}
\emailAdd{chethan.krishnan@gmail.com}
\emailAdd{pnbalasubramanian@gmail.com }
\affiliation[a]{International Center for Theoretical Sciences, \\
Indian Institute of Science Campus, \\
  Bangalore - 560012, \ India}
\affiliation[b]{Center for High Energy Physics, \\
  Indian Institute of Science, \\ 
  Bangalore - 560012, \ India}
\abstract{We argued in arXiv:1408.0624 that the quartic scalar field in AdS has features that could be instructive for answering the gravitational stability question of AdS. Indeed, the conserved charges identified there have recently been observed in the full gravity theory as well. In this paper, we continue our investigation of the scalar field in AdS and provide evidence that in the Two-Time Formalism (TTF), even for initial conditions that are far from quasi-periodicity, the energy in the higher modes at late times is exponentially suppressed in the mode number. Based on this and some related observations, we argue that there is no thermalization in the scalar TTF model within time-scales that go as $\sim 1/\epsilon^2$, where $\epsilon$ measures the initial amplitude (with only low-lying modes excited). It is tempting to speculate that the result holds also for AdS collapse.
}

\setcounter{tocdepth}{2}
\maketitle

\section{Overview}\label{introduction}

The question of whether AdS space is stable \cite{Bizon, BizonReview} against turbulent thermalization and the formation of black holes under generic (non-linear) perturbations has received much attention recently. AdS space with conventional boundary conditions is like a box, and therefore perturbations that were weak to begin with can reflect multiple times from the boundary, potentially resulting in sufficient localization of energy to create black holes. Aside from the fact that black hole formation is a question of fundamental interest in (quantum) gravity, this problem acquires another interesting facet via the AdS/CFT correspondence: it captures the physics of thermalization in strongly coupled quantum field theories. 

At the moment however, it is fair to say that the evidence for and against the instability of AdS when excited by low-lying, low-amplitude modes is mixed \cite{Leo, Buchel, CEH1, BKS1, CEH2, Buchel2, I-Sheng, Freivogel, Bizon2}. In an effort to (partially) clarify this situation, in this paper we will make some comments about two loosely inter-related questions:
\begin{itemize}
\item Does ``most" initial data lead to thermalization?  

\item Can one argue that within a time-scale of order ${\cal O}(1/\epsilon^2)$, where $\epsilon$ captures the amplitude of the initial pertubration, thermalization does (not?) happen? This is an interesting question because the statement of \cite{Bizon} is that black hole formation happens in AdS within this time scale. 
\end{itemize}
We will ask these questions, which are inspired by gravitational (in)stability in AdS, in the context of a simpler problem: a self-interacting $\phi^4$ scalar field in AdS. The works of \cite{BKS1, Freivogel, I-Sheng, CEH2, Buchel2} suggest that these systems have close similarities, so we believe that this effort will be instructive and worthwhile. 

One of our main tools will be the Two-Time Formalism (TTF) developed in \cite{Buchel} (we will describe this approach in section 2). We will argue why this approach has various advantages, and why we believe it captures the essential physics of resonances in the full (ie., non-TTF) model. But we emphasize that this will shed light on the instability question only if the instability, if it exits, is caused by resonances (which seems plausible to us). If the instability is caused by some other (possibly longer time-scale) dynamics, TTF in the leading order can miss that physics. 
But we expect that physics in the ${\cal O}(1/\epsilon^2)$ time-scale should be captured by TTF.

Furthermore, for concreteness, we will take the following as the definition of the {\em absence of thermalization}: the presence of exponentially distributed energies in the higher modes, as a function of the mode number\footnote{Note that the definition of thermalization is somewhat ambiguous. We are adopting this as a {\em sufficient} but {\em not necessary} condition for the absence of thermalization as we will make more precise at the beginning of Section \ref{results}. One source of ambiguity is that our system is classical and suffers from a UV catastrophe: so once the system has fully thermalized, the average energy per state would be zero, if we don't truncate it.  In particular, the distribution of energies should {\em not} be compared to a canonical ensemble distribution, rather it should be thought of as capturing the efficiency of energy transfer to higher modes.}. That is, if the system has  $A_j \sim e^{-j \beta}$ at late times for some positive $\beta$ we will say that it does not thermalize (at least for a very long time). Loosely, one could also adopt a definition that the system has instability towards thermalization if the late-time behavior of the $j$'th mode goes as $A_j \sim j^{-\alpha}$, where $\alpha$ is a positive quantity -- it is possible however that this is not a necessary nor a sufficient condition \cite{Buchel2}, and we will not use this in our paper.

Within the context of these three limitations (namely, working with the scalar field and in the TTF approach and within a particular definition of thermalization) our results imply the following answers for the two questions (combined into one):

\begin{itemize}
\item Initial data with only low lying modes do {\em not} lead to thermalization for the quartic scalar field in the TTF formalism within a time-scale of  ${\cal O}(1/\epsilon^2)$. This suggests that if at all there is thermalization in the full theory, it should be coming from non-resonant transfer of energy.



\end{itemize} 

Together, we believe that these observations present fairly strong evidence that thermalization (as defined above) does not happen for initial value data which have only the low-lying modes excited. Our results, as already emphasized, are for the $\phi^4$-scalar: but we believe 
 similar statements apply for AdS gravity as well. We make various further comments of varying degrees of technicaility in later sections.
 
This result might seem at odds with the numerical results for the Gaussian initial value profile discussed in \cite{Bizon}. But it has been pointed out in \cite{Buchel2} that the 
spectrum of the Gaussian profile in AdS, in fact has energy ditributed in the higher modes as a power law to begin with. So it is not a suitable initial initial profile to resolve between the presence or absence of collapse using our criterion for the absence of thermalization, which requires exponential suppression in the higher modes. Clearly more work is required to clarify whether this is an acceptable initial profile or not for resolving this particular question. 


For completeness, lets also state that our results are still not quite conclusive. Apart from the points emphasized above, there is also the perverse possibility that collapse happens, but not due to resonances -- but note however that the time-scale for this will be bigger than $\sim 1/\epsilon^2$.

\section{TTF Formalism}\label{ttf}
The action for the scalar field theory is given by 
\begin{align}
S = \int d^{x}x \sqrt{-g} \, \left(\dfrac{1}{2}\nabla_{\mu}\phi \nabla^{\mu}\phi + V(\phi)\right)
\end{align}
where the potential is given by 
\begin{align}
V(\phi) = \dfrac{\lambda}{4!} \phi^{4}
\end{align}
The metric for the space is given by
\begin{align}
ds^{2} = \sec^{2}x\left(-dt^{2} + dx^{2} + \sin^{2}x\, d\Omega^{2}\right)
\end{align}
The equations of motion for the scalar field are given by
\begin{align}\label{scalareom}
\phi^{(2,0)} + \square_{s}\phi \equiv \phi^{(2,0)} - \phi^{(0,2)} -\dfrac{2}{\sin x\,\cos x}\phi^{(0,1)} = -\dfrac{\lambda}{6 \cos^{2}x} \phi^{3}
\end{align}
where the $ \square_{s} $ represents the spatial Laplacian operator. This operator has an eigenfunction basis given by
\begin{align}
\square_{s} e_{j}(x) &= \omega^{2}_{j} \,e_{j}(x)\\
e_{j}(x) &= 4\sqrt{\dfrac{(j+1)(j+2)}{\pi}} \cos^{3} x\,\, _{2}F_{1}\left(-j,j+3;\dfrac{3}{2}; \sin^{2}x\right) \label{mode}\\
\omega_{j}^{2} &= (2 j +3)^{2}\quad j=0,1,2,\dots \label{omega}
\end{align}
The inner product in this basis is defined as 
\begin{align}
 \left(f,g\right) = \int dx\,\tan^{2}x \,f(x) \, g(x)
\end{align}
In the \emph{Two-Time Framework} (TTF), we have the slow-moving time defined as $ \tau = \eps^{2}t $, which requires the time derivatives to be redifined as $ \partial_{t} \rightarrow \partial_{t} + \eps^{2}\partial_{\tau} $. The scalar field is written as an expansion in the small-parameter $ \eps $ as
\begin{align} 
\phi = \eps\, \phin{1} + \eps^{3}\phin{3} + \mathcal{O}(\eps^{5}) 
\end{align} 
Note that the ratio between the slow and fast times ($\tau$ and $t$) also controls the overall scale of the amplitude.
Putting this expansion in the scalar field equations of motion eq.(\ref{scalareom}) we get
\begin{align} 
\text{order $ \eps $:}& \qquad \partial_{t}^{2}\phin{1} - \partial_{x}^{2}\phin{1} -\dfrac{2}{\sin x \cos x} \partial_{x}\phin{1} = 0\\
\text{order $ \eps^{3} $:}&\qquad\partial_{t}^{2}\phin{3} + 2\partial_{t}\partial_{\tau}\phin{1} - \partial_{x}^{2}\phin{3} \nonumber\\
&\qquad\qquad\qquad -\dfrac{2}{\sin x \cos x} \partial_{x}\phin{3} = -\dfrac{\lambda}{6\cos^{2}x}\phi^{3}_{(1)}(t,\tau,x) 
\end{align} 
The order $ \eps $ equation has the general real solution
\begin{align} 
\phin{1} = \sum_{j=0}^{\infty} \left(A_{j}(\tau) e^{-i \omega_{j}t} +\overline{A}_{j}(\tau) e^{i \omega_{j}t} \right) e_{j}(x) \label{sol}
\end{align} 
Note that the introduction of the slow times gives an extra variable that we can tune - we will use this at order $ \eps^{3} $ to cancel of the resonant terms. The equations that accomplish this are called the TTF equations. 
Substituting the above first order results into the order $ \eps^{3} $ equations we get
\begin{align} 
&\hspace{-1cm}\partial_{t}^{2}\phin{3} -2 i  \sum_{k=0}^{\infty}\omega_{k} \left(\partial_{\tau}A_{j}(\tau) e^{-i \omega_{j}t} - \partial_{\tau}\overline{A}_{j}(\tau) e^{i \omega_{j}t} \right)e_{j}(x) + \square_{s}\phin{3} \nonumber\\
&= -\dfrac{\lambda}{6\cos^{2}x}\sum_{j,k,l=0}^{\infty} \Biggl[\left(A_{j}(\tau) e^{-i \omega_{j}t} +\overline{A}_{j}(\tau) e^{i \omega_{j}t} \right)\left(A_{k}(\tau) e^{-i \omega_{k}t} +\overline{A}_{k}(\tau) e^{i \omega_{k}t} \right) \nonumber\\
&\hspace{4cm}\qquad\left(A_{l}(\tau) e^{-i \omega_{l}t} +\overline{A}_{l}(\tau) e^{i \omega_{l}t} \right) e_{j}(x) e_{k}(x)e_{l}(x)\Biggr]
\end{align} 
Projecting on the basis solutions give
\begin{align} 
&\left(e_{j}(x), [\partial_{t}^{2}+\omega_{j}^{2}]\phin{3}\right) - 2i \omega_{j}\left[\partial_{\tau}A_{j}(\tau) e^{-i \omega_{j}t} - \partial_{\tau}\overline{A}_{j}(\tau) e^{i \omega_{j}t} \right] \nonumber\\
&\hspace{2cm}= -\dfrac{\lambda}{6} \sum_{k,l,m=0}^{\infty}C_{jklm}\Biggl[ \left[A_{k}(\tau) e^{-i \omega_{k}t} +\overline{A}_{k}(\tau) e^{i \omega_{k}t}\right]\left[A_{l}(\tau) e^{-i \omega_{l}t} +\overline{A}_{l}(\tau) e^{i \omega_{l}t} \right] \nonumber\\
&\hspace{6cm}\left[A_{m}(\tau) e^{-i \omega_{m}t} +\overline{A}_{m}(\tau) e^{i \omega_{m}t} \right]\Biggr]
\end{align} 
where 
\bea
C_{jklm} = \int_{0}^{\pi/2} dx \tan^{2}(x) \sec^{2}(x) \,\,e_{j}(x) e_{k}(x) e_{l}(x) e_{m}(x) \label{cijkl}
\eea
By direct computation (using properties of Jacobi polynomials -- which are an alternate way to describe the basis functions, see Appendix A), one can show that the necessary and sufficient condition for resonances is 
\bea
\omega_j+\omega_m=\omega_k+\omega_l.
\eea
The absence of other combinations for the resonances for the scalar theory was recognized and used in \cite{BKS1} (see footnote 3 of \cite{CEH2} for a simple proof). They are also absent in the gravity case, but the computation required to show this in that case is substantially more lengthy \cite{CEH1}. The close parallel between the structure of the resonances in the two cases is evidently one of the reasons why they exhibit similarities in their thermalization dynamics \cite{BKS1}.

In any event, at this stage we have the freedom to choose the $A_j(\tau)$ as mentioned above so that the resonances on both sides are cancelled. This is accomplished by solving the $A_j$ according to
\bea
-2 i\omega_j\  \partial_\tau A_j=-\frac{\lambda}{6} \sum_{k,l,m=0}^{\infty}C_{jklm} A_k A_l {\bar A_m} \label{TTF}
\eea
and its complex conjugate. These are the {\em TTF equations} that we will use extensively in the next section. Once the resonances are cancelled, the coupling to the higher modes is expected to be weak and we believe it is unlikely that there will be efficient channels for thermalization: but this is a prejudice, and possibly far from proof. In any event, we can systematically solve for $\phin{3}$ at this stage if we wish, without being bothered by resonances.

Note that the simplicity of the quartic scalar arises from the fact that the $C_{ijkl}$ have a (relatively) simple expression. We will comment more about this in Appendix A.

Before concluding this section we quote some pertinent results from \cite{BKS1} for our scalar TTF system. Firstly, we can get the TTF equations using an effective Lagrangian
\begin{align}
L_{TTF}=i \sum_i \omega_i ( B_i \dot{\bar{B_i}}- \bar B_i \dot{ B_i})+\sum  C_{ijkl} \bar{ B}_i(\tau) \bar{ B}_j(\tau)  B_k(\tau)  B_l(\tau) ,
\label{ttf-lag}
\end{align}
where summation in the interaction term is over $i, j, k, l$ such that $\omega_{i}+\omega_{j}-\omega_{k}-\omega_{l}=0$. In writing the expression in this form, we have done an appropriate scaling of each mode by $\omega_k$ and $\lambda$ for easy comparison with the notation of \cite{BKS1}: $B_k$ are the rescaled modes.  The system has a dilatation symmetry: $B_k(\tau) \rightarrow \epsilon B_k(\frac{1}{\epsilon^2} \tau)$. So if thermalization happens in the TTF theory it should scale inversely as the square of the amplitude: the assumption that TTF theory captures the relevant physics is the assumption that the system {\em has} such a scaling regime. 

However, the system has the following conserved charges \cite{BKS1} arising from a corresponding set of symmetries:
\bea
Q_0=\sum  B_k \bar{ B}_k, \ {\rm symmetry:} \ B_k \rightarrow e^{i\theta}  B_k , \hspace{1.2in} 
\\
Q_1=\sum k  B_k \bar{ B}_k,\ {\rm symmetry:} \  B_k \rightarrow e^{i k \theta}  B_k ,\hspace{1.1in} \\
E=\sum_{\omega_i+\omega_j-\omega_k-\omega_l=0}  C_{ijkl} \bar{ B}_i(\tau) \bar{  B}_j(\tau)  B_k(\tau)  B_l(\tau), \ {\rm Symmetry:}\  t \rightarrow t+\alpha.
\eea
Various pieces of evidence indicating that the evolution of the quartic scalar in AdS has some close connections to collapse in AdS gravity were presented in \cite{BKS1}. The above conserved charges were identified for the full gravity system in \cite{CEH2} (see also \cite{Buchel2}).

\section{Results}\label{results}

In this section, we will study various aspects of the TTF equations for the quartic scalar in some detail. As mentioned in the introduction, we will take the exponential decay of $A_j(\tau)$ with $j$ as an indication that thermalization is suppressed. In \cite{Buchel2} some arguments were made that a power law $A_j \sim j^{-a}$ for positive $a$ is indicative of thermalization/black hole formation. We will make this somewhat more precise as follows. The basic object that is taken as an indicator of collapse in \cite{Bizon, Buchel} is the quantity $|\Pi(t,0)|^2$, the unbounded growth of whose profile is taken as the onset of collapse. The analogue of this quantity in our scalar TTF case can be taken as $|\dot \phi_{(1)}(t,0)|^2$ (compare Figure 1(A) and the accompanying discussion in \cite{BKS1} to Figure 3 in \cite{Buchel}). At this point, using (\ref{sol}), (\ref{mode}) and (\ref{omega}) we can see that this quantity can be estimated and bounded via
\bea
|\dot \phi_{(1)}(t,0)|^2 \sim |\sum j^2 A_j|^2 \lesssim \sum j^2|A_j|^2.
\eea
Now, it is evident that when $A_j \sim e^{-j \alpha}$ the last quantity is finite and therefore the LHS can never diverge, which is what we set out to show. This indicates that expoential suppression of higher modes is a sufficient condition for {\em absence} of thermalization. Note however that we are silent about what constitutes thermalization at the level of modes -- fortunately, we will never need a precise definition of that for the purposes of this paper.

\begin{figure}\begin{center}
\includegraphics[scale=0.5]{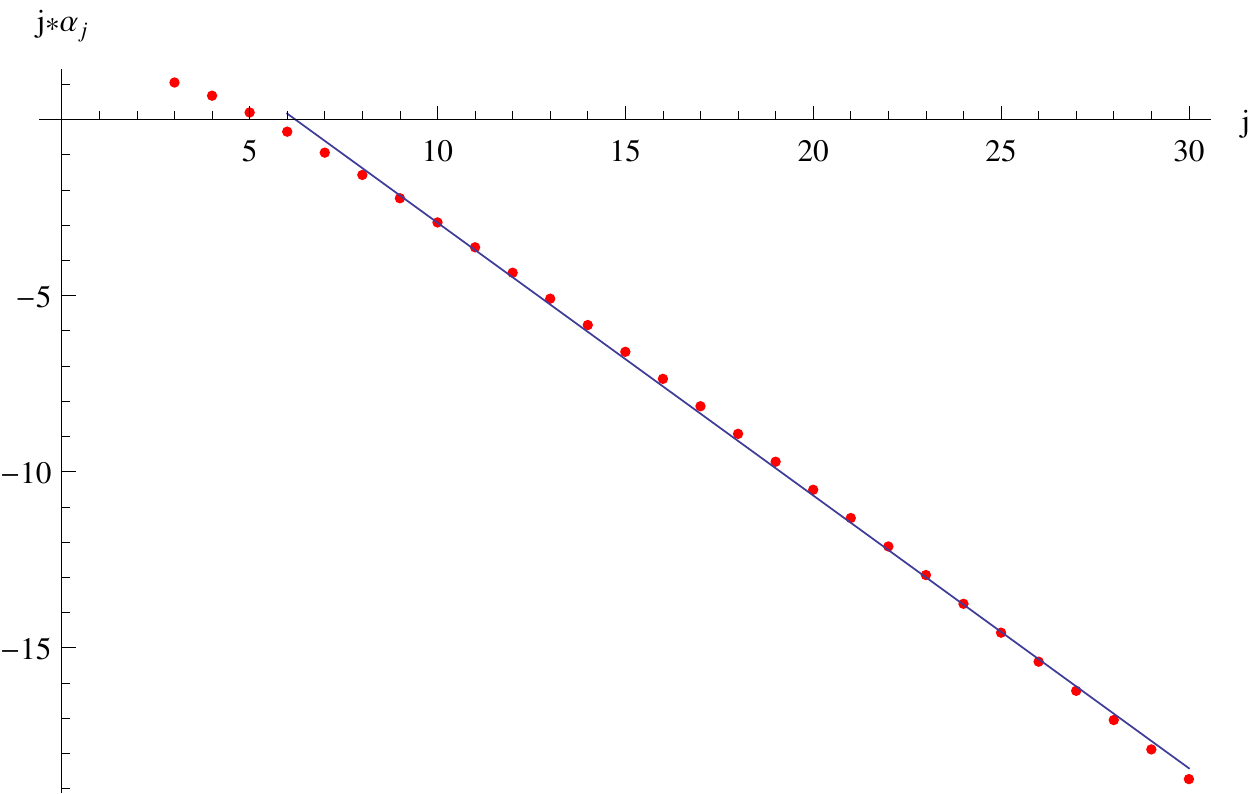}
\caption{The log-plot of $j \alpha_j$ vs. $j$ for the quasi-periodic solutions. The linear fit is indicative of exponential suppression of $A_j$ with $j$.}\label{quasidecay}\end{center}
\end{figure}

The TTF theory has quasi-periodic solutions (see \cite{Buchel} for a discussion of analogous solutions in the gravity system) of the form
\bea 
A_j(\tau) =\alpha_j \exp(-i\beta_j \tau), \ {\rm where} \  \beta_j=\beta_0+j (\beta_1-\beta_0).
\eea
One can choose $\alpha_0,\alpha_1$ (or $\beta_0,\beta_1$) and determine the rest of the $\alpha_j$ via the TTF equations (\ref{TTF})\footnote{For some initial conditions we see more than one quasiperiodic solution.}, if one truncates the system at some $j=j_{max}$ and demand stability of the solution against variation in $j_{max}$. We have done this, and the resulting modes decay with the mode number $j$ as $\sim \frac{\exp(-c j)}{j}$ for some positive $c$, see Figure (\ref{quasidecay}). This is obviously consistent with our definition of (non-)thermalization. In \cite{Buchel} the $j_{max}$ was taken to be $\sim 50$, in our case we are able to go up to $j_{max}=150$. 

\begin{figure}
\begin{center}
\begin{subfigure}[figA]{0.4\textwidth}
\includegraphics[scale=0.4]{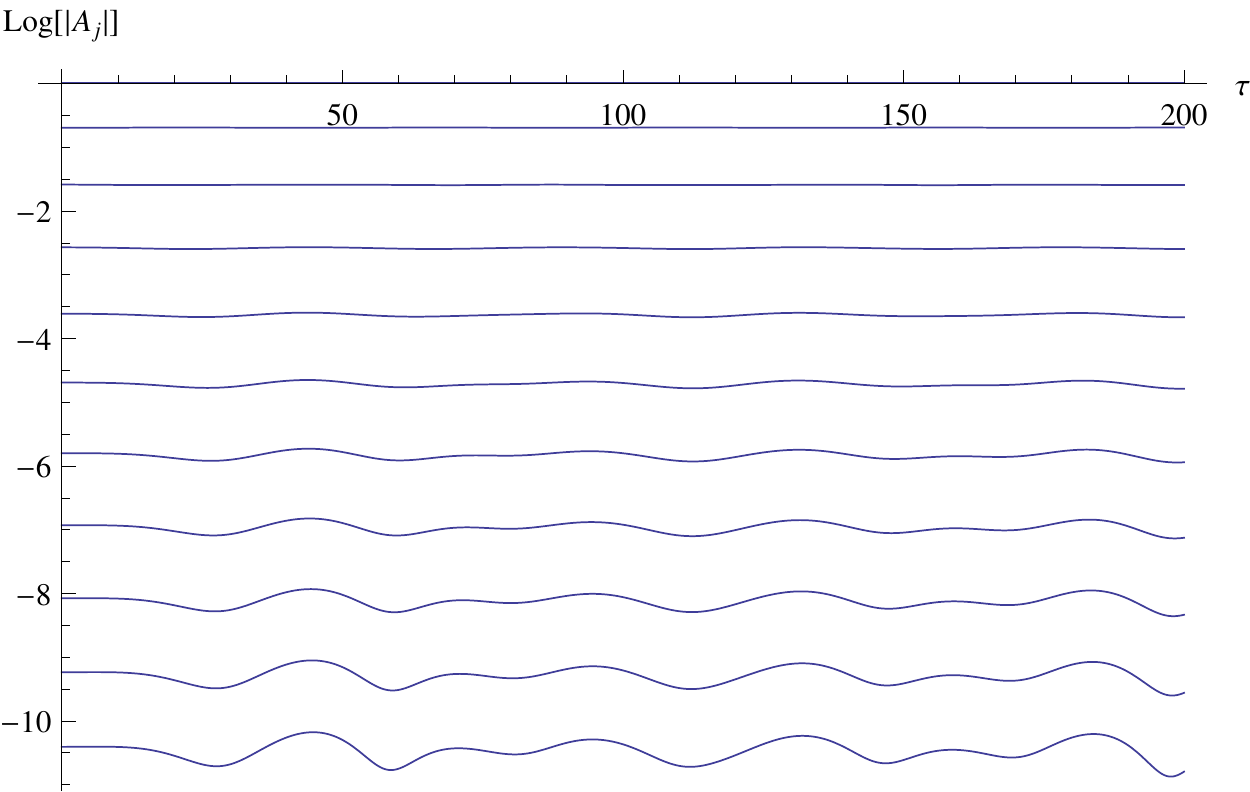}
\caption{Representative plots of ``small" perturbations around quasi-periodic solutions.}\label{weakpertplot}
\end{subfigure}
\begin{subfigure}[figA]{0.4\textwidth}
\includegraphics[scale=0.4]{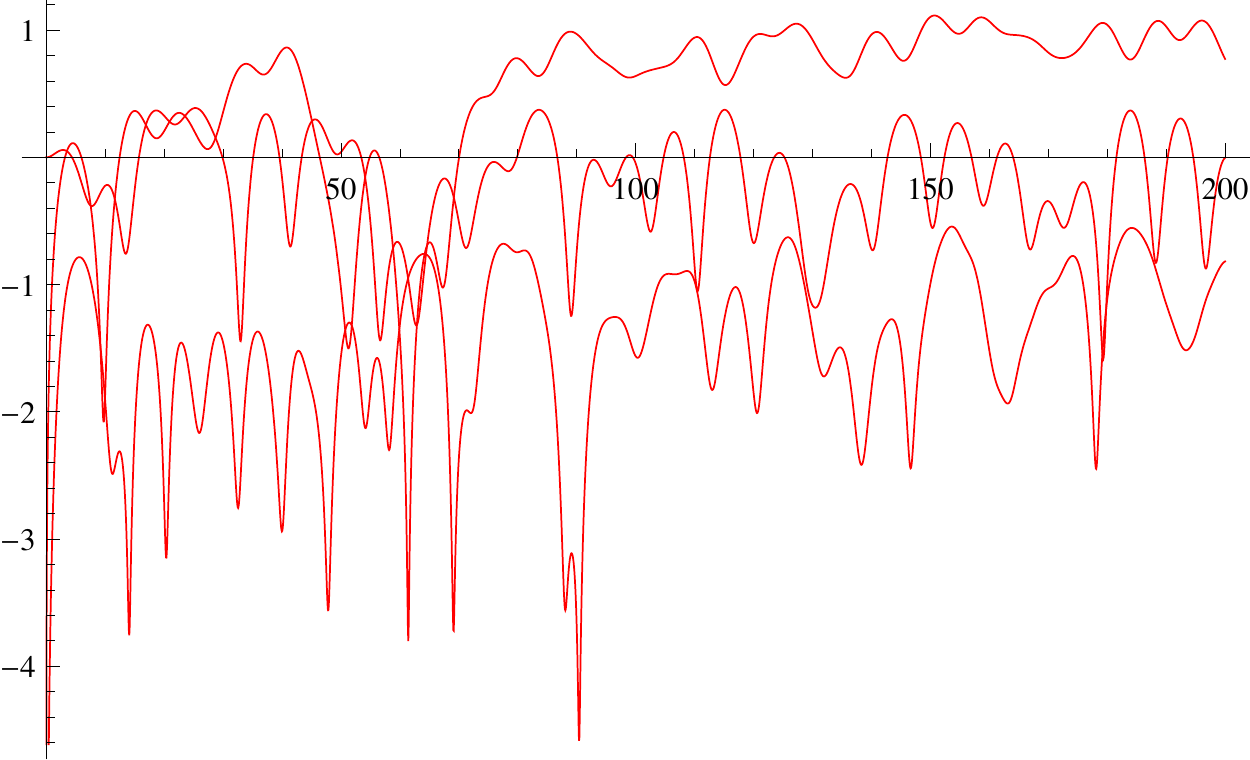}
\caption{Representative plots of ``large" perturbations around quasi-periodic solutions.}\label{strongpertplot}
\end{subfigure}
\caption{Evolution plots of perturbations around quasi-periodic solutions.}\label{pertplots}\end{center}
\end{figure}

If we perturb a quasi periodic solution we expect to get oscillations  of the $A_j$'s around $\alpha_j$. See Figure (\ref{weakpertplot}) for solutions where the initial value of the $A_j$ are close\footnote{In order to make these statements precise, we will need a notion of closeness between solutions in terms of modes. A convenient way to define a dimensionless measure of the ``distance" between two solutions (say 1 and 2) is to consider
\bea
\Delta_{12}
=\frac{\sum_j A_j^{(1)} A_j^{(2)*}}
{\sqrt{\sum_k |A^{(1)}_k|^2}\sqrt{\sum_l |A^{(2)}_l|^2}}\label{norm}
\eea 
$\Delta_{12} \sim 1$ is close. The summation is only up to mode number $j_{max}$. 
}
to their quasiperiodic values. If on the other hand, the initial $A_j$ values are sufficiently far from their quasiperiodic values, we expect that the solutions transition to chaos. This expectation is qualitatively verified in Figure (\ref{strongpertplot}) where we launch the $A_j$ far away from quasi-periodicity.
In what follows we will show that {\em even} in these far-from quasiperiodic solutions, the maximum value attained by the $A_j$ as we evolve the solution is exponentially suppressed in $j$. This is an indication that energy transfer to the higher modes is suppressed even in these solutions - if this behavior holds also in gravity, it could be an indication that these solutions generically do not collapse.



\begin{figure}
\begin{center}
\begin{subfigure}[figA]{0.4\textwidth}
\includegraphics[scale=0.4]{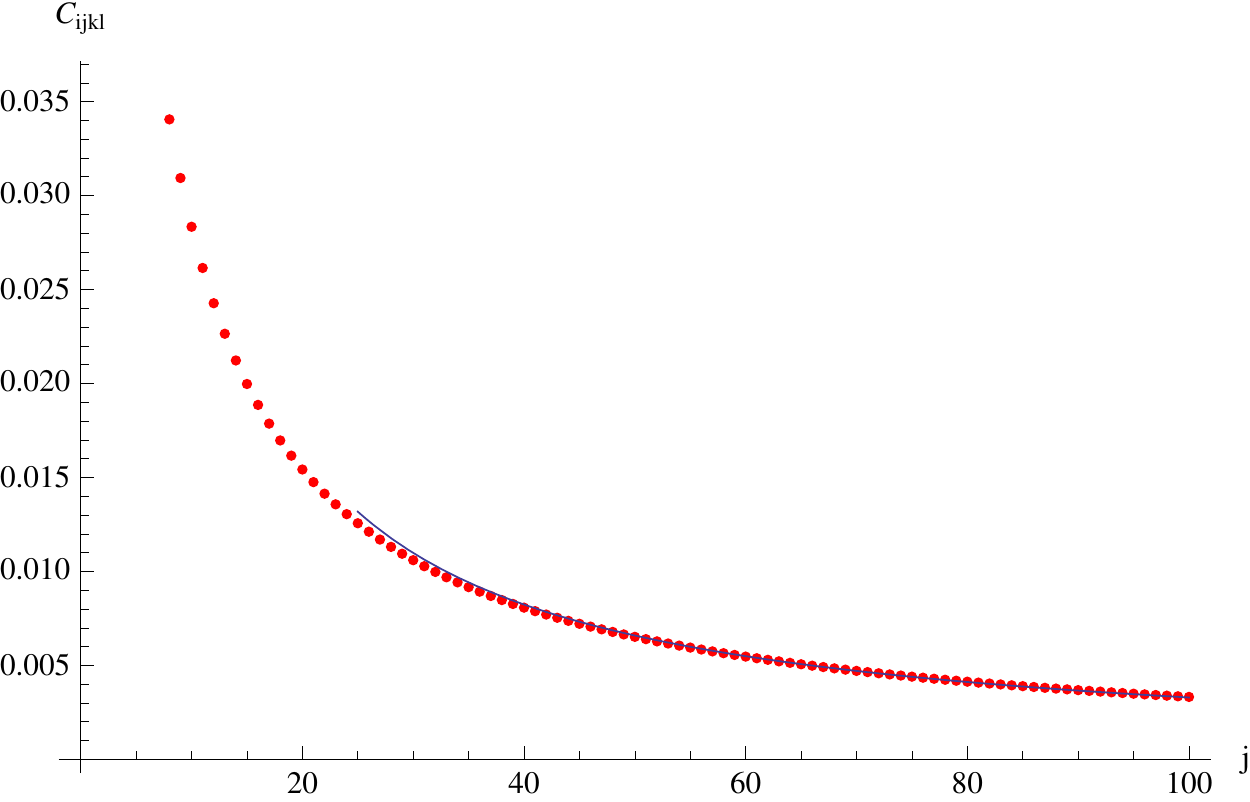}
\caption{$i,j,k,l =1,2,n,n+1$.}
\end{subfigure}
\begin{subfigure}[figA]{0.4\textwidth}
\includegraphics[scale=0.4]{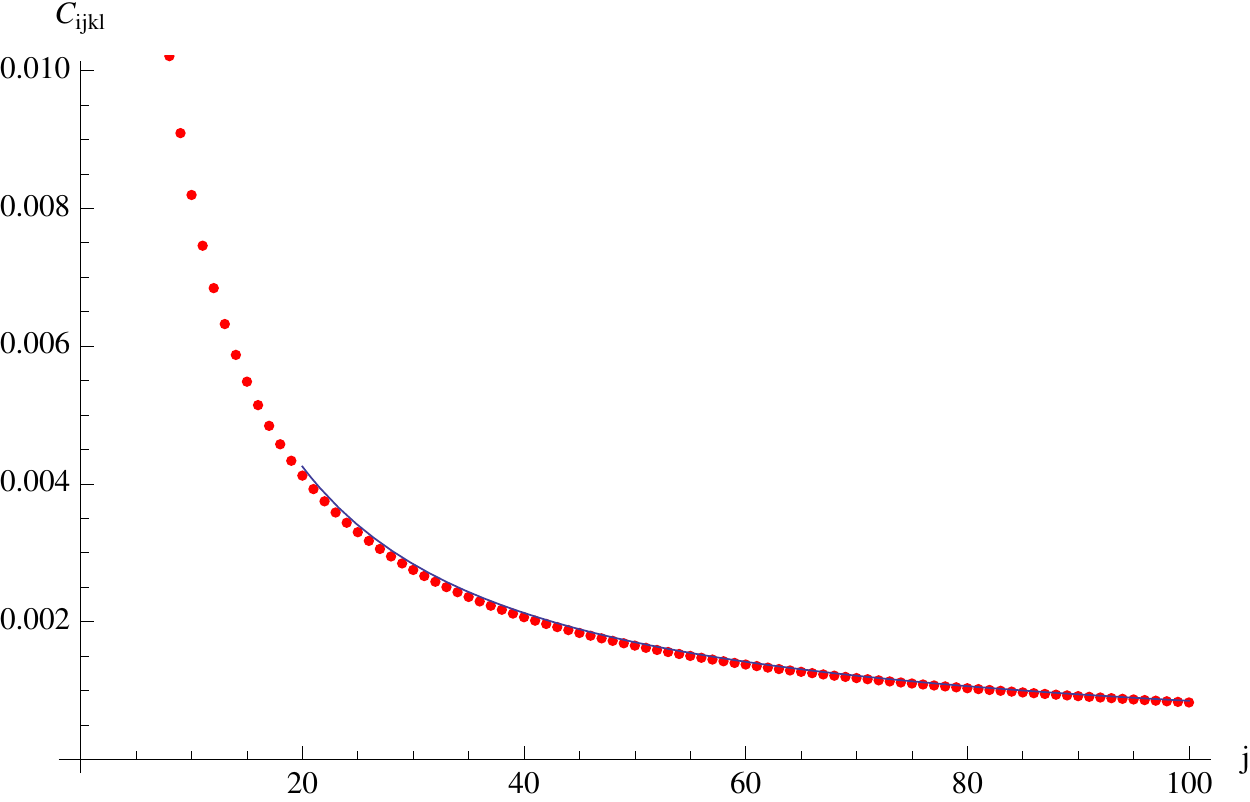}
\caption{ $i,j,k,l =n, 2n, 3n, 4n$.}\label{cijklplot}
\end{subfigure}
\caption{Plot of $C_{ijkl}$ together with an inverse linear fit.}\label{Cplot}
\end{center}
\end{figure}

One of the ways in which one might try to understand the efficiency of energy transfer to higher modes is by studying the coefficients\footnote{Note that the coefficients $C_{ijkl}$ can be determined via (\ref{cijkl}) analytically, but using Mathematica. Some comments on this are given in Appendix A.} $C_{ijkl}$ which signify the coupling between the modes. To understand the behaviour of TTF equations at large $j$, we look at various kinds of limits we may consider for $C_{ijkl}$ as the $i, j, k, l$ are sent to $\infty$. One is  a simple scaling of indices, ${i, j, k, l} \rightarrow {a i, a j, a k, a l}$. By fitting the plot (see Figure \ref{cijklplot}), we see that in this case $C_{ijkl}$ goes as ${\cal O}(\frac{1}{a})$ as $a \rightarrow \infty$.
Another case is where we keep two modes fixed and take another two to infinity:   $i \sim j \sim {\rm approximately \ fixed}$, but with $k \sim l \sim a$ and we take $a \rightarrow \infty$. We find that they also have a ${\cal O}(\frac{1}{a})$ fall off. It is important to note that because of the resonance condition, these are the only possible couplings available for a high mode - one cannot (for example) hold three indices small while sending the forth one to infinity. So progressively higher modes are weakly coupled, both to each other as well as to the low-lying modes. 

\begin{figure}
\begin{center}
\includegraphics[scale=0.4]{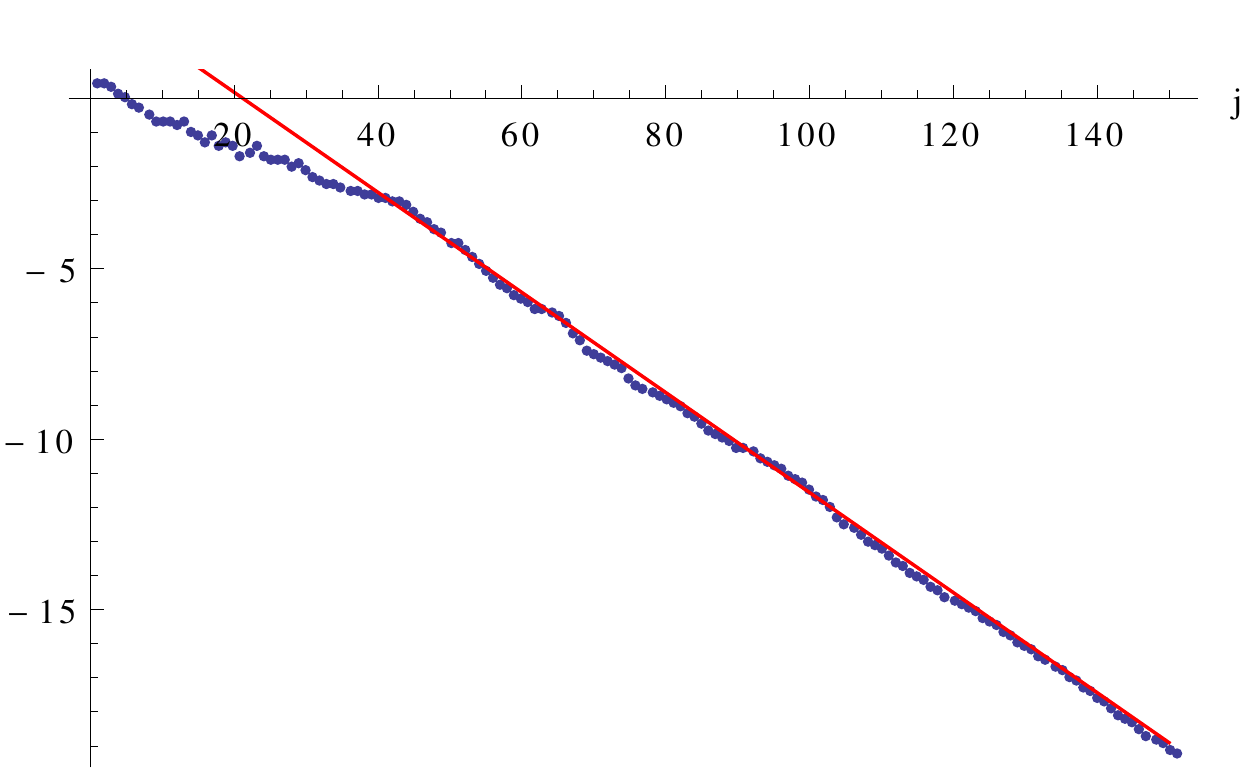}
\subcaption{$A_0=A_1=A_2=A_3=A_4=1$}
\includegraphics[scale=0.4]{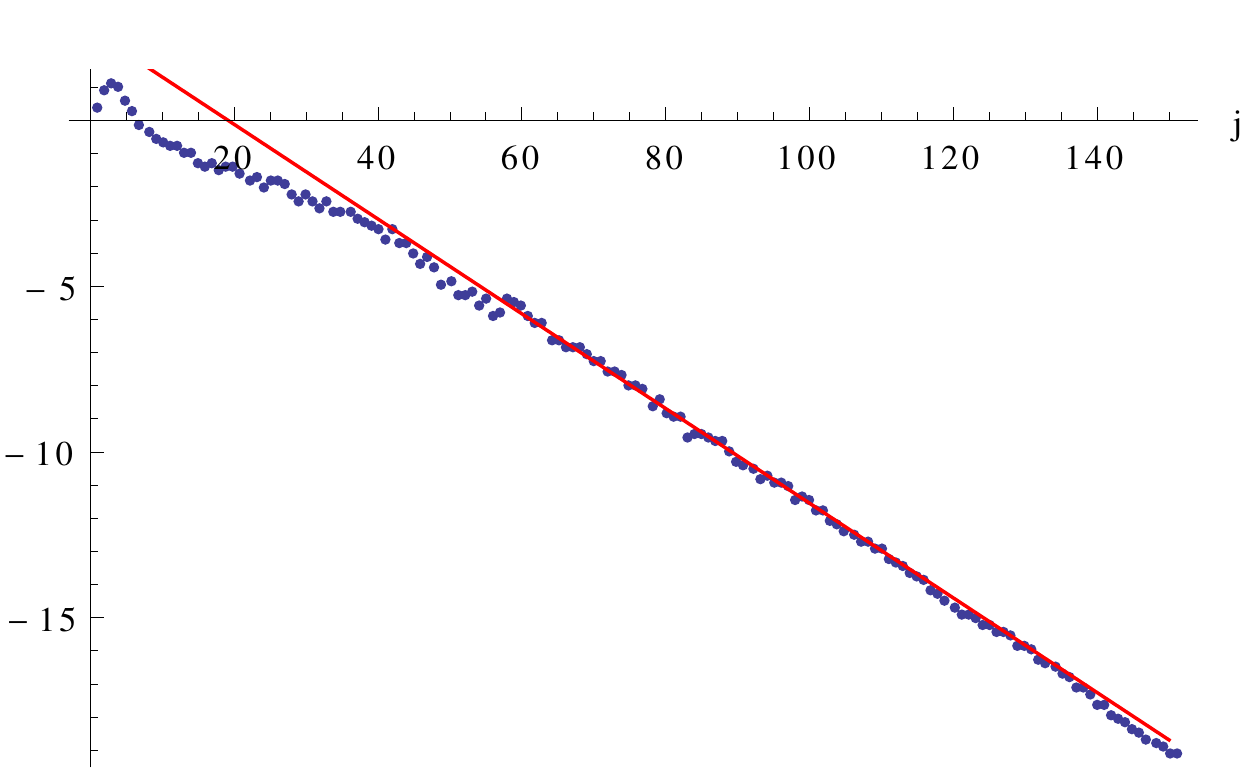}
\subcaption{$A_0=1,A_1=i,A_3=3$}
\includegraphics[scale=0.4]{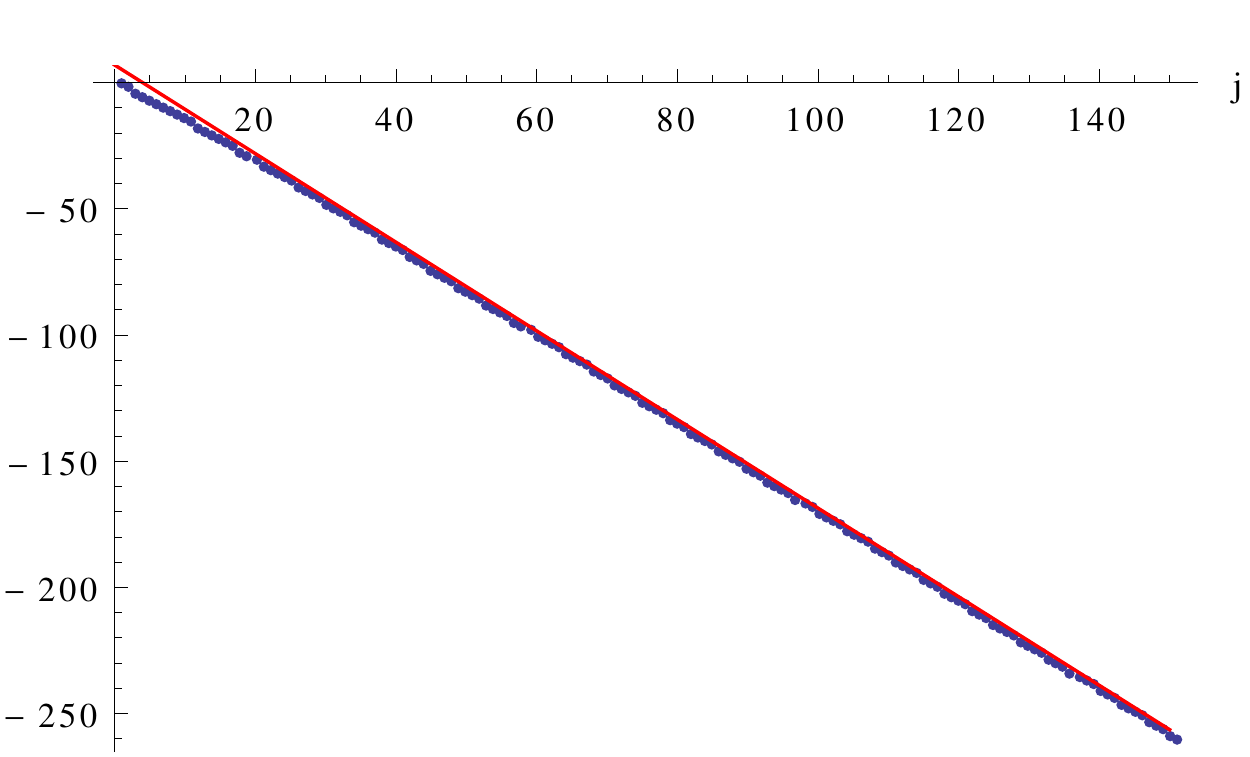}
\subcaption{$A_0=1,A_1=0.1$}
\label{expfall}
\end{center}
\caption{Plot of $\log[{\rm Max}[A_n(\tau)]]$ for $j\in [0,150]$ and a linear fit. The fit has been done in the region $j\in [40,150]$. The last figure corresponds to quasi-periodic initial data.}\label{lastplot}
\end{figure}

Finally, we consider the evolution of the modes when we launch the system both near and far from quasi-periodic initial conditions. The way we do this is by calculating the coefficients $C_{ijkl}$ analytically (see Appendix A for some comments on this) and then integrating the resulting TTF equations numerically for the various initial conditions. In all cases we plot maximum value of $A_j$ that is attained during the entire period of evolution 
against $j$, and we find that this ${\rm Max}[A_j(\tau)]$ exponentially decays with $j$ for {\em all initial data}. We see an exponential decay with respect to $j$, not just for solutions close to quasi-periodic solutions, but also for those that are far from it: see Figure (\ref{lastplot}). This is true even though for some initial conditions we see a approximately power law decay of modes up to some intermediate frequency. These statements can be verified using the norm (\ref{norm}) with the understanding that the summation over $j$ has to be restricted to be above some appropriately chosen $j_{min}$ (and of course below $j_{max}$) when we are talking about high modes. 

 
\acknowledgments
We thank Oleg Evnin for discussions, and Piotr Bizon and Andrzej Rostworowski for correspondence.\\

\appendix
\section{Comments on $C_{ijkl}$ and Jacobi Ploynomials}

The determination of the $C_{ijkl}$ is in principle straightforward by direct evaluation of (\ref{cijkl}). This is an analytically tractable problem because the basis functions $e_j(x)$ can be written in terms of Jacobi polnomials as
\bea
e_j(x)=4\sqrt{\dfrac{(j+1)(j+2)}{\pi}}\  \frac{\Gamma(j+1)\Gamma(3/2)}{\Gamma(j+3/2)}\ \cos^3 x\ P^{(1/2,3/2)}_j(\cos 2x).
\eea
Jacobi polynomials are (orthogonal) polynomials in their arguments and therefore in our case they merely involve only (a finite number of) powers of sinusoids\footnote{Explicit expressions for Jacobi polnomials can be found in numerous places ranging from Wikipedia to Abramowitz and Stegun.}. Therefore the integral for $C_{ijkl}$, which is in the range $[0, \pi/2]$ can, again in principle, be straightforwardly evaluated. It turns out that the result can be expressed in terms of finite sums of finite products of Gamma functions and such, but simplifying them on Mathematica becomes time-consuming. One could in principle try to simplify the expressions manually, but we have adopted a more pragmatic approach: we  evaluate the integrals analytically on Mathematica by re-expressing the powers of sinusoids in terms of product formulas. Since the integrals are over $[0, \pi/2]$ this makes them substantially less intensive as far as time requirements are considered. This way we are able to algorithmize the (analytic) computation of $C_{ijkl}$ on Mathematica, after which we use them in the TTF equations to do our numerical evolutions.


\bibliographystyle{JHEP}

\end{document}